 \documentclass[%preprintnumbers,%preprint,prd,a4paper,
twocolumn,showpacs, amsmath, amssymb, floatfix,
nofootinbib,wrapfloat byrevtex]{revtex4}

 \usepackage{amsmath,mathrsfs,graphicx,epstopdf,placeins,float}
 \usepackage{booktabs}
 \usepackage{multirow}
\usepackage{color}

 \newcommand{\beq}[1]{\begin{equation}\label{#1}}
 \newcommand{\eeq}{\end{equation}}
 \newcommand{\bea}[1]{\begin{eqnarray}\label{#1}}
 \newcommand{\eea}{\end{eqnarray}}
 \newcommand\figcaption{\def\@captype{figure}\caption}
 \newcommand\tabcaption{\def\@captype{table}\caption}

\begin{document}

 \title{Two particle entanglement and its geometric duals}
 \author{M. Abdul Wasay $^{\!\!1,2}$}
\email{wasay31@gmail.com}

\author{Asma Bashir$^{1}$}
\affiliation{$^1$Department of Physics, University of Agriculture, Faisalabad 38040, Pakistan.\\
$^2$National Centre for Physics, Quaid-i-Azam University Campus, Islamabad, Pakistan.}
 \begin{abstract}
 We show that for a system of two entangled particles, there is a dual description to the particle equations in terms of classical theory of conformally stretched spacetime. We also connect these entangled particle equations with Finsler geometry. We show that this duality translates strongly coupled quantum equations in the pilot-wave limit to weakly coupled geometric equations.
  \end{abstract}
 %\pacs{04.20.Cv,~03.65.Ta}

 \maketitle
 \smallskip

\section{Introduction}

 The forces of nature are described by fields which one defines on the spacetime except gravity, which is defined by the spacetime itself as explained by general relativity. In general relativity, physical effects are described elegantly in terms of differential geometry of curved spacetime. All the interactions in GR are completely described by purely geometric equations. In short, GR says that gravity is nothing but a curvature in the spacetime fabric caused by heavy physical objects placed on it. Thus GR, as it stands today, is one of the most successful theories of modern physics

 Quantum mechanics however is concerned with probability interpretation of objects and phenomenon which is not very clear and intuitive as compared to mathematically beautiful GR.
Thus it naturally led many physicists to attempt to reformulate quantum mechanics in a geometric language, like GR.  Ref. \cite{a}-\cite{Alon2} are a few such efforts towards the geometrical rewriting of quantum laws.

 It is sometimes useful to develop different mathematical theories describing the same physics. Such an equivalent description of different theories is very helpful at times to avoid certain difficulties; for instance, the well-known AdS/CFT duality
 \cite{ADS} provides an equivalence between a specific gravitational theory and a lower dimensional non-gravitational theory, and is helpful in answering some difficult questions arising on one side of the correspondence by manipulations on the other side. Specifically, the duality is the following
 \bea{}
 String~ Theory~ on ~ AdS_5\times S^5\sim~~~~~~~~~~~~~~~~~~~~~~~~~~~\nonumber
 \\
 \mathcal{N}=4
 ~ SU(N)~ gauge~ theory~ in~ 4D \nonumber
 \eea
 There are also other dualities in string theory which interconnect various string theories and untangle difficulties related to either side.

 Concerning a similar approach, various results have been reported on supersymmetric quantum mechanical models and their topological aspects, a few of which are \cite{SQM1,SQM2,SQM3,SQM4,SQM5}, since the introduction of a topological index by Witten \cite{SQM6}.

In order to reformulate quantum mechanics in a geometric fashion, one needs to associate physical reality to objects and to define background space.
Quantum correlation rests almost entirely on the consideration of non-locality between spatially separated particles. To write  quantum mechanics in a geometric way, it is thus important to include determinism in the quantum description.
%\textcolor[rgb]{1.00,0.00,0.00}{One could have a well-defined theory by specifying trajectories through the guiding equation of the form $\textcolor[rgb]{1.00,0.00,0.00}{m_i}\frac{dx_i}{dt}=\nabla_i S$. This guiding equation together with Schr\"{o}dinger equation then describe the dynamics of the system.}
When the wave function is factored using the ansatz \cite{26,27} $\psi=Pe^\frac{iS}{\hbar}$ into amplitude and phase, the particle momentum can be expressed in terms of the guiding equation $m_i\frac{dx_i}{dt}=\nabla_i S$ and
writing the Schr\"{o}dinger equation using this ansatz leads to the following equation of motion

\bea{}
m_i\frac{d v_i}{dt}=-\nabla_i(V_i+Q)
\label{a}
\eea
$P$ and $S$ represent the amplitude and dimensionless phase of the pilot wave respectively.
The additional term is the "quantum potential".
%The interpretation of quantum mechanics by means of hidden variables introduces the "quantum potential",
\bea{}
Q=\sum_{i=1}^n\frac{\hbar^2}{2m_i}\frac{\nabla_i^2|\psi|}{|\psi|}
\label{q}
\eea
which includes the position of all the constituent particles. Keeping \eqref{q} in mind, one can see from \eqref{a} that dynamics of single particle is specified by the entire system and thus non-locality is inherent in the system. It is actually this term which associates physical reality with the particles.

In \cite{24} it was shown following the formalism of \cite{26,27} that how relativistic Klein-Gordon equation can be rephrased in a geometric way and how the particle trajectories can be represented by geodesics on a conformally curved $4n$ dimensional configuration space. A similar approach \cite{28} dealt with the non-relativistic limit of the above.

The curious and bizarre phenomenon of quantum entanglement has attracted physicists since long. Its principle importance is due to its implications for quantum information processing. As mentioned above, there are some works reporting a correspondence between geometry and physics, in view of these it seems plausible to find geometric alternatives for quantum entanglement. The purpose of this paper is to present an alternate (geometric) language for quantum entanglement in the pilot wave limit. We will  examine a simple case of two particle entanglement in terms of physical trajectories and write the purely physical equations of quantum entanglement by using the factored form of the wave function described above and then translate these physical equations to geometry in two ways, 1: in terms of $1+6$ dimensional conformally rescaled configuration space geometry, 2: in terms of Finsler geometry. As a reference model, we will consider a system of two identical particles (with spin) in the vicinity of external magnetic field described by the spin incorporated Schr\"{o}dinger equation. Nonlocality is present in the system through the quantum potential. Choice of the particular model is well-suited to envision how external magnetic field assists particle's spin to be entangled.

The paper is organized as: Section-II is devoted to the model description and derivation of the physics side of four entanglement equations in the pilot-wave limit. In section-III, we present the (configuration space) geometric interpretation of quantum entanglement equations from section-II. In section-IV, we extend the Einstein-Hilbert action for Finsler spacetime and show that particle equations from section-II can also be written in the language of Finsler geometry. The results are briefly summarized in section-V.
\section{Two particle entanglement}
Two particles are said to be entangled if the quantum states of two particles could not be described independently, even if the two particles are at cosmic spatial distance. The state of two entangled particles can not be separated as product states
\bea{}
\psi(x_1,x_2,t)\not=\psi(x_1,t)\psi(x_2,t)\nonumber
\eea

Since if one of the particle is in a particular state of spin 'up' the the other particle must have spin 'down'. Now
we need to incorporate spin of the particles, the spin incorporated Schr\"{o}dinger equation is given by

\bea{}
i\hbar\frac{\partial\psi}{\partial t}\!\!=\!\!\left(\!-\frac{\hbar^2}{2m_j}\nabla^2_j-\frac{\hbar^2}{2m_j}\frac{e^2_j}{c^2\hbar^2}A_j^2(x_j)-\frac{\mu_j.B(x_j)\textbf{S}_j}{S_j}\!\right)\!\!\psi
\label{1}
\eea

where, $e_j$ is the charge of the particles, $\mu_j$ is the magnetic moment due to magnetic affects produced by the spin of the particle, $B(x_j)$ is the magnetic field, $A_j(x_j)$ is the magnetic vector potential due to magnetic field $B(x_j)$, $\textbf{S}_j$ is the spin operator and $S_j$ its eigenvalue: the spin of the particle.
The index $j$ represents which one of the two particles is affected.
There is no spin-spin coupling because the particles may be spatially separated to large distances and there will be no significant interaction between the magnetic fields generated by the two particles. But the magnetic field of the two particles do have interaction with the applied external magnetic field and they will experience a torque. For this we need to consider the last term  which accounts for the potential energy $\mu.B(x_j)$ of the magnetic moment in the external magnetic field. In double index notation one can rewrite Eq.\eqref{1} as

\bea{}
\sum\limits_j[\frac{\hbar^2}{2m_j}\partial^K_j\partial_{jK}+\frac{e^2_j}{2m_jc^2}A_j^{K2}(x_j)+\frac{\mu_j.B(x_j)\textbf{S}_j}{S_j}\nonumber
\\
+i\hbar\partial_0]\psi=0
\label{g}
\eea
where, $j=1,2$ and $K$ is the space index in $3$ dimensions.
\subsection{First equation}
Following \cite{26,27},  the wave function can be factored into amplitude and phase as

\bea{}
\psi(x_1,x_2,t)=P(x_1,x_2,t)exp\left(\frac{iS(x_1,x_2,t)}{\hbar}\right)
\label{2}
\eea

 Using \eqref{2} in \eqref{g} leads to

\bea{}
\sum\limits_j\left(\frac{\hbar^2}{2m_j}\partial^K_j\partial_{jK}+\frac{e^2_j}{2m_j}A_j^{K2}(x_j)+\frac{\mu_j.B(x_j)\textbf{S}_j}{S_j}+i\hbar\partial_0\right)\nonumber
\\
\times P(x_1,x_2,t)\textmd{exp}\left(\frac{iS(x_1,x_2,t)}{\hbar}\right)=0
\label{06}
\eea

We work in the limit of absolute "time", designated as $t$. In this limit, space and time are not on equal footing so the
spatial and temporal derivatives will be treated differently.
Moreover, $\partial P/\partial t=0$, as for large $t$ the amplitude on the average is zero. We get (see appendix A)
%\bea{}
%\sum\limits_j\left(\frac{\hbar^2}{2m_j}\partial^K_j\partial_{jK}Pe^\frac{iS}{\hbar}+\frac{i\hbar^2}{2m_j\hbar}P\partial^K_j(\partial_{jK}S)e^\frac{iS}{\hbar}+\frac{e^2_j}{2m_j}A_j^{K2}(x_j)Pe^\frac{iS}{\hbar}
%+\frac{\mu_j.B(x_j)\textbf{S}_j}{S_j}Pe^\frac{iS}{\hbar}-P\partial_0S\right)=0\nonumber
%\eea
%\bea{}
%\sum\limits_j\left(\frac{\hbar^2}{2m_j}\partial^K_j\partial_{jK}P+\frac{i\hbar^2}{2m_j\hbar}P\partial^K_j(\partial_{jK}S)\frac{iS}{\hbar}+\frac{e^2_j}{2m_j}A_j^{K2}(x_j)P
%+\frac{\mu_j.B(x_j)\textbf{S}_j}{S_j}P-P\partial_0S\right)=0\nonumber
%\eea
%\bea{}
%\sum\limits_j\frac{\hbar^2}{2m_j}\frac{\partial^K_j\partial_{jK}P}{P}=\sum\limits_j\left(\frac{(\partial^K_jS)(\partial_{jK}S)}{2m_j}-\frac{e^2_j}{2m_j}A_j^{K2}(x_j)-\frac{\mu_j.B(x_j)\textbf{S}_j}{S_j}+\partial_0S\right)\nonumber
%\eea
\bea{}
{2m_j}Q(x_1,x_2,t)=\sum\limits_j[(\partial^K_jS)(\partial_{jK}S)-e^2_jA_j^{K2}(x_j)\nonumber
\\
-{2m_j}\frac{\mu_j.B(x_j)\textbf{S}_j}{S_j}+{2m_j}\partial_0S]
\label{3}
\eea

On the L.H.S of Eq.\eqref{3} above, we have used the definition $Q=\sum\limits_j\frac{\hbar^2}{2m_j}\frac{\partial_j^K\partial_{jK}P(t,\vec{x_j})}{P(t,\vec{x_j})}$.

This is the first equation of two particle entanglement. Here $Q(x_1,x_2,t)$ is the quantum potential and $P(x_1,x_2,t)$ is the associated pilot wave. If entanglement is lost, the quantum potential will be the sum of two terms with each term depending on the position of one particle only. Since the quantum potential locks the position of two particles, if one of the two particles has its magnetic moment aligned to the external magnetic field, the other particle will have an anti-aligned magnetic moment. The two particles will then experience torque in the opposite directions (one clockwise and the other counterclockwise). Therefore, we need to consider this specific form of Schr\"{o}dinger equation to understand how the spins of two particles are entangled.
\\
\subsection{Second equation}
 %In quantum mechanics, the probability density is given by
%\bea{}
%\rho(r)=|\psi(r)|^2=\psi^\ast\psi\nonumber
%\eea
%\bea{}
%\frac{\partial}{\partial t}(\psi^\ast\psi)=\psi^\ast\frac{\partial\psi}{\partial t}+\frac{\partial\psi^\ast}{\partial t}\psi\nonumber
%\eea
The wave function, as defined above, permits to construct the conserved current as
\bea{}
\partial_0(\psi^*\psi)-\sum\limits_{j=1}^n\partial_j^m\left(\frac{i\hbar}{2m_j}(\psi^\ast\overleftrightarrow{\partial}_{jm}\psi)\right)=0
\label{l}
\eea

Using the factored form of the wavefunction from Eq.\eqref{2} in Eq.\eqref{l} leads to

\bea{}
\partial_0(P^2)+\sum\limits_j\partial_{jK}\left(\frac{P^2}{m_j}(\partial_j^K S)\right)=0
\label{4}
\eea

 Eq.\eqref{4} is the second of the entanglement equations, representing the conserved current.

\subsection{Third equation}

In order to characterize the particle trajectories in terms of the pilot wave, one needs to define a guiding equation as follows

\bea{}
\frac{dx_j^K}{dt}=\frac{\hbar}{m_j}~\textmd{Im}\!\left(\frac{\psi,D_j^K\psi}{\psi,\psi}\right)(x_1,x_2)
\label{h}
\eea

with the following definition of the covariant derivative

\bea{}
D_j^K\psi=\nabla_jPe^\frac{iS}{\hbar}-\frac{ie_j}{c\hbar}A_j^K(x_j)Pe^\frac{iS}{\hbar}
\eea
%\bea{}
%D_j^K\psi=\frac{i}{\hbar}(\partial_j^KS)Pe^\frac{iS}{\hbar}-\frac{ie_j}{c\hbar}A_j^K(x_j)Pe^\frac{iS}{\hbar}\nonumber
%\eea
%\bea{}
%\frac{dx_j^K}{dt}=\frac{\hbar}{m_j}Im\left(\frac{\frac{i}{\hbar}(\partial_j^KS)Pe^\frac{-iS}{\hbar}Pe^\frac{iS}{\hbar}-\frac{ie_j}{\hbar}A_j^K(x_j)Pe^\frac{-iS}{\hbar}Pe^\frac{iS}{\hbar}}
%{Pe^\frac{-iS}{\hbar}Pe^\frac{iS}{\hbar}}\right)\nonumber
%\eea
%\bea{}
%\frac{dx_j^K}{dt}=\frac{1}{m_j}Im\left(\frac{i(\partial_j^KS)P^2-ie_jA_j^K(x_j)P^2}
%{P^2}\right)\nonumber
%\eea

and using $\psi$ from Eq.\eqref{2} in Eq.\eqref{h}, we get

\bea{}
\frac{dx_j^K}{dt}=\frac{(\partial_j^KS)-ie_jA_j^K(x_j)}{m_j}
\eea

Momentum is given by

\bea{}
\pi_j^K=(\partial_j^KS)-ie_jA_j^K(x_j)
\label{5}
\eea

Each of the two particles follow a trajectory specified by the other particle (described by the same wave function). In fact, the wave function can geographically stretch over the entire universe. This mysterious interdependence of wave function interlocks both particles into single physical reality.

\subsection{Fourth equation}

The equation of motion for two entangled particles is given by (see appendix B)

\bea{}
m_j\frac{d^2x_j^K}{dt^2}=\partial_{lN}\left[2 Q+\frac{2\mu_j.B(x_j)\textbf{S}_j}{S_j}-\frac{ieSF^*}{m_j}-2\partial_0S\right]
\label{6}
\eea
where,
\bea{}
F^*=\partial_j^K A_l^N+\partial_l^N A_j^K
\eea
So, the particle trajectory is specified by the quantum potential  $Q(x_1,x_2,t)$ and combined electric and magnetic effects $[\frac{2\mu_j.B(x_j)\textbf{S}_j}{S_j}-\frac{ieSF^*}{m_j}]$ and is guided by the pilot wave.

%\bea{}
%m_j^2\frac{d^2x_j^K}{dt^2}=\sum\limits_l\left(\partial_{lN}(\partial_l^NS)(\partial_j^KS)-\partial_{lN}(\partial_j^KS)ie_lA_l^N(x_l)
%-\partial_{lN}(\partial_l^NS)ie_jA_j^K(x_j)+\partial_{lN}ie_lA_l^N(x_l)ie_jA_j^K(x_j)\right)\nonumber
%\eea
%\bea{}
%m_j^2\frac{d^2x_j^K}{dt^2}=\sum\limits_l\left(\partial_{lN}(\partial_l^NS)\left[(\partial_j^KS)-ie_jA_j^K(x_j)\right]-\partial_{lN}ie_lA_l^N(x_l)\left[(\partial_j^KS)-ie_jA_j^K(x_j)\right]\right)\nonumber
%\eea
%\bea{}
%m_j^2\frac{d^2x_j^K}{dt^2}=\sum\limits_l\partial_{lN}(\partial_l^NS)\pi_j^K-\partial_{lN}ie_lA_l^N(x_i)\pi_j^K\nonumber
%\eea
This equation confirms that the two particles are entangled because the motion of each particle is effected by the quantum potential which depends on the position of both particles.
This makes entanglement in this approach even more perceptible, because due to nonlocality, the particles are so coupled that the state of one particle is entirely specified by the other particle.

In Eqs.\eqref{3},\eqref{4},\eqref{5},\eqref{6} $P$, $S$ and $Q$ of the two particles depend on $1+6$ dimensions, $6$ space dimensions with a single time coordinate. One can define

\bea{}
x^L=(t,\vec{x_1}^1,\vec{x_1}^2,\vec{x_1}^3,\vec{x_2}^1,\vec{x_2}^2,\vec{x_2}^3)
\eea

such that $\partial_j^K,\partial_l^N\rightarrow\partial^L$ and $\partial_{jK},\partial_{lN}\rightarrow\partial_L$; with

\bea{}
Q=\frac{\hbar^2}{2m_j}\frac{\partial^L\partial_LP(t,\vec{x_j})}{P(t,\vec{x_j})}.\nonumber
\eea

All four entanglement equations obtained above can now be written as

\bea{}
{2m_j}Q(x_1,x_2,t)=[(\partial^LS)(\partial_{L}S)-e^2_jA^{L2}(x_j)\nonumber
\\
-{2m_j}\frac{\mu_j.B(x_j)\textbf{S}_j}{S_j}+{2m_j}\partial_0S]
\label{7}
\eea

\bea{}
\partial_L\left(\frac{P^2(\partial^L S)}{m_j}\right)+\partial_0(P^2)=0,
\label{8}
\eea
\bea{}
\pi^L=(\partial^LS)-ie_jA^L(x_j),
\label{9}
\eea
\bea{}
\hspace{-2mm} m_j\frac{d^2x^L}{dt^2}\!=\!\partial_{L}\!\!\left[2Q\!+\!\frac{2\mu_j.B(x_j)\textbf{S}_j}{S_j}-2\partial_0S\!-\!\frac{ieSF^*}{m_j}\right]
\label{10}
\eea
Note that in contrast to \cite{24}, this model is concerned with spin-$1/2$ particles and is non-relativistic. In (relativistic model) \cite{24}, the geometric theory was developed in $4n$ dimensions but in our case it is necessary to work with $6$-dimensions of space and one of time to picture two particle entanglement. In non-relativistic limit, $\psi^*\psi$ can be interpreted as probability density, but it is not possible to provide an interpretation for the probability of Klein-Gordon equation by $\psi^*\psi$. The probability interpretation of Klein-Gordon equation is given in terms of Klein-Gordon current, that is conserved with respect to time.
The effects of external field interaction are absorbed into the field strength tensor defined as
\bea{}
F=\partial_j^K A_l^N-\partial_l^N A_j^K \nonumber
\eea
but here $\partial_j^K$ is not a four gradient as in \cite{24} and the tensor defines the electromagnetic field in three dimensional space.
In the following section, we will show that this set of four equations can be written in a geometric way.

\section{Entanglement equations in terms of 1+6 dimensional CONFIGURATION SPACE}
We consider a $1+6$ dimensional configuration space of $2$-particles with a single time coordinate. The coordinates are defined as

\bea{}
\hat{x}^\Lambda=(\hat{t},\hat{\vec{x}}_1^1,\hat{\vec{x}}_1^2,\hat{\vec{x}}_1^3,\hat{\vec{x}}_2^1,\hat{\vec{x}}_2^2,\hat{\vec{x}}_2^3)
\eea

The scalar equation specifying curvature in this setup employs $1+6$ dimensions, and is given by

\bea{}
P_s\left(\hat{R}+k\hat{L}_M\right)=\hat{R}+k\hat{L}_M
\label{11}
\eea

Here $P_s$ is symmetrization operator between different particles $x_i^\lambda$ and $x_j^\lambda$, $\hat{R}$ is Ricci scalar, $\hat{L}_M$ is matter Lagrangian and $k$ is coupling constant that accounts for the matter-field interaction and now with this symmetrization condition, the particle action reads

\bea{}
S\left(\hat{g}_{\Lambda\Delta}\right)=\int dt\int dx^{6}\sqrt{\left|\hat{g}\right|}\left(\hat{R}+k\hat{L}_M\right)
\label{12}
\eea
The metric $\hat{g}$ is factorized into a conformal function $\phi(\vec{x}_j,t)$and a flat part $\eta$ \cite{24,28} to describe local conformal part of the theory. The conformal transformation here is given by
\bea{}
\hat{g}_{\Lambda\Gamma}=\phi^{\frac{4}{5}}\eta_{LG}
\label{13}
\eea

This rescaling however does not change the Physics.
The inverse of the metric is given by
\bea{}
\hat{g}^{\Lambda\Gamma}=\phi^{\frac{-4}{5}}\eta^{LG}
\label{14}
\eea
The lower Greek and lower Roman index are identified as $\hat{\partial}_\Lambda=\partial_L$ so that the adjoint derivatives are different in each notation i.e.,
\bea{}
\hat{\partial}^\Lambda=g^{\Lambda\Sigma}\hat{\partial}_\Sigma=\phi^{\frac{-4}{5}}\eta^{LS}\partial_S\nonumber
\\
\hat{\partial}^\Lambda=\phi^{\frac{-4}{5}}\partial^L\nonumber
\\
\hat{\partial}_\Lambda=\phi^{\frac{4}{5}}\partial_L\nonumber
\eea
\subsection{Geometric dual to the first equation:}

The particle action in terms of $\phi$ and $g_{LD}$ is

\bea{}
S\left(\phi,g_{LD}\right)=-\int dt\int dx^{6}\sqrt{|g|}~[\frac{24}{5}(\partial^L\phi)(\partial_L\phi)+\nonumber
\\
\phi^2(R+k(L_M-\frac{\mu_j.B(x_j)\textbf{S}_j}{S_j}))]\nonumber
\eea

In this model we are concerned with flat Minkowski background space so $g_{LG}=\eta_{LG}$ and $|g|=1$ and $R=0$. So the action simplifies to
\bea{}
S\left(\phi\right)=-\int dt\int dx^{6}[\frac{24}{5}(\partial^L\phi)(\partial_L\phi)+\phi^2k(L_M\nonumber
\\
-\frac{\mu_j.B(x_j)\textbf{S}_j}{S_j})]\nonumber
\eea

The equation of motion is
\bea{}
-\frac{24}{5}\frac{\partial^L\partial_L\phi}{\phi}=k\left(L_M-\frac{\mu_j.B(x_j)\textbf{S}_j}{S_j}\right)
\label{15}
\eea
The matter Lagrangian $L_M$ is given by
%\bea{}
%L_M=\frac{2\hat{\pi}^\Lambda\hat{\pi}_\Lambda}{2\hat{M}_G}+\frac{\partial S_H(t,\vec{x}_j)}{\partial t}\nonumber
%\eea
\bea{}
L_M=\frac{2(\hat{\partial}^\Lambda S_H-ie_j\hat{A}^\Lambda(x_j))(\hat{\partial}_\Lambda S_H-ie_j\hat{A}_\Lambda(x_j))}{2\hat{M}_G}\nonumber
\\
+\frac{\partial S_H}{\partial t}\nonumber
\eea
The equation of motion with this Lagrangian then gives
\bea{}
-{2\hat{M}_G}\frac{24}{k(5)}\frac{\partial^L\partial_L\phi}{\phi}=2(\hat{\partial}^\Lambda S_H)(\hat{\partial}_\Lambda S_H)-2e^2_j\hat{A}^{\Lambda2}(x_j)\nonumber
\\
-{2\hat{M}_G}\frac{\mu_j.B(x_j)\textbf{S}_j}{S_j}+{2\hat{M}_G}\dot{S}_H
\label{16}
\eea
With the following matching conditions
\bea{}
k=-\frac{24}{5}.\frac{2\hat{M}_G}{\hbar^2}\nonumber
\eea
\bea{}
\phi(\vec{x}_j,t)=P(\vec{x}_j,t)\nonumber
\eea
\bea{}
S_H(\vec{x}_j,t)=S(\vec{x}_j,t)\nonumber
\eea
\bea{}
m_j=\hat{M}_G\nonumber
\eea
we find that Eq.\eqref{16} is identical to Eq.\eqref{7}.
\subsection{Geometric dual to the second equation:}
%The stress energy tensor is given by
%%\bea{}
%%T^{\Lambda\Delta}=2\frac{\delta L_M}{\delta g_{\Lambda\Delta}}+g^{\Lambda\Delta}L_M\nonumber
%%\eea
%%\bea{}
%%T^{\Lambda\Delta}=2\frac{\delta}{\delta g_{\Lambda\Delta}}\left(g_{\Lambda\Delta}\frac{(\hat{\partial}^\Lambda S_H)(\hat{\partial}^\Delta S_H)}{\hat{M}_G}+\partial_0S_H-\frac{e_j^2\hat{A}^{\Lambda2}}{\hat{M}_G}\right)\nonumber\\+
%%\frac{g^{\Lambda\Delta}(\hat{\partial}^\Lambda S_H)(\hat{\partial}_\Lambda S_H)}{\hat{M}_G}+g^{\Lambda\Delta}\partial_0S_H-g^{\Lambda\Delta}\frac{e_j^2\hat{A}^{\Lambda2}}{\hat{M}_G}\nonumber
%%\eea
%\bea{}
%T^{\Lambda\Delta}=\frac{2(\hat{\partial}^\Lambda S_H)(\hat{\partial}^\Delta S_H)}{\hat{M}_G}+~~~~~~~~~~~~~~~~~~~~~~~~\nonumber
%\\
%g^{\Lambda\Delta}\left(\frac{(\hat{\partial}^\Lambda S_H)(\hat{\partial}_\Lambda S_H)}{\hat{M}_G}+
%\partial_0S_H-\frac{e_j^2\hat{A}^{\Lambda2}}{\hat{M}_G}-\textcolor[rgb]{1.00,0.00,0.00}{\frac{\mu_j.B(x_j)\textbf{S}_j}{S_j}}\right)\nonumber
%\eea
%
%where $\partial_0=\partial/\partial t$ and since the stress energy tensor is covariantly conserved, so
%
%\bea{}
%\nabla_\Lambda T^{\Lambda\Delta}=0\nonumber
%\eea
%
%From this, we can write
%
%\bea{}
%\frac{(\hat{\partial}^\Delta S_H)\nabla_\Lambda(\hat{\partial}^\Lambda S_H)}{\hat{M}_G}=0
%\label{17}
%\eea
%\bea{}
%\frac{(\hat{\partial}^\Lambda S_H)\nabla^\Delta(\hat{\partial}_\Lambda S_H)}{\hat{M}_G}=0
%\label{18}
%\eea
%\bea{}
%\frac{(\hat{\partial}_\Lambda S_H)\nabla_\Lambda(\hat{\partial}^\Lambda S_H)}{\hat{M}_G}+\nabla_\Lambda(\partial_0 S_H)=0
%\label{19}
%\eea
Using conservation of energy-momentum tensor
\bea{}
\nabla_\Lambda T^{\Lambda\Delta}=0
\eea

We find (see appendix C)
\bea{}
\frac{(\hat{\partial}_\Lambda S_H)\nabla_\Lambda(\hat{\partial}^\Lambda S_H)}{\hat{M}_G}+\nabla_\Lambda(\partial_0 S_H)=0
\label{19}
\eea

The Levi-Civita connection is given by
\bea{}
\Gamma^\Sigma_{\Lambda\Delta}=\frac{1}{2}g^{\Sigma\Xi}\left(\partial_\Lambda g_{\Delta\Xi}+\partial_\Delta g_{\Xi\Lambda}-\partial_\Xi g_{\Lambda\Delta}\right)
\label{j}
\eea
Making use of the conformal rescaling of the metric defined above, Eq.\eqref{j} leads to
\bea{}
\Gamma^\Sigma_{\Lambda\Delta}=\frac{1}{2}\phi^{\frac{-4}{5}}[(\partial_L\phi^{\frac{4}{5}})\delta^S_D+(\partial_D\phi^{\frac{4}{5}})\delta^S_L
-(\partial^S\phi^{\frac{4}{5}})\eta_{LD}]
\eea
With this, the relation in \eqref{19} reads:
\\
For the first term in \eqref{19}
\bea{}
\nabla_\Lambda(\hat{\partial}^\Lambda S_H)=
\phi^{\frac{-14}{5}}\partial_L(\phi^2(\partial^LS_H))=0
\label{20}
\eea

For the second term in \eqref{19}
\bea{}
\nabla_\Lambda(\partial_0S_H)=\partial_0(\partial_LS_H)
\label{21}
\eea

From \eqref{19}, with first term and second term as above, we get

\bea{}
\left(\frac{\partial_L\left[\phi^2(\partial^LS_H)\right]}{\hat{M}_G}+\partial_0(\phi^2)\right)=0
\label{22}
\eea

With the matching conditions defined above, Eq.\eqref{22} is identical to Eq.\eqref{8}.
Note that $\phi(x,t)$ enters the theory in two ways, $\phi^2(x,t)$ gives probability interpretation and $\phi(x,t)$ accounts for matter and field interaction.

\subsection{Geometric dual to the third equation:}

The momenta is defined by the derivative of $S_H$ (Hamilton principle function) as suggested by Hamilton-Jacobi formalism.
\bea{}
\hat{\pi}^\Lambda=(\hat{\partial}^\Lambda S_H-ie_j\hat{A}^\Lambda(x_j))
\label{23}
\eea

Which is identical to the third equation Eq.\eqref{9}.

\subsection{Geometric dual to equation of motion:}
The geometric dual to the trajectory equation of motion Eq.\eqref{10}, is the following (see appendix D),

\bea{}
\hat{M}_G\frac{d^2\hat{x}^\Lambda}{d\hat{s}^2}=\hat{\partial}_\Delta\left[-\frac{24}{k(5)}Q
+\frac{\mu_j.B(x_j)\textbf{S}_j}{S_j}-\frac{ieS_H\hat{F}_{\Lambda}^*}{\hat{M}_G}-\dot{S}_H
\right]
\label{24}
\eea
where,
\bea{}
\hat{F}_{\Lambda}^*=(\hat{\partial}_\Delta \hat{A}_\Lambda+
\hat{\partial}_\Lambda \hat{A}_\Delta)\nonumber
\eea
%For instantaneous information processing, we may choose $\partial_0S=0$.

where $s=t$. We can see that Eq.\eqref{24}, with the defined matching conditions, is identical to Eq.\eqref{10}.
%One can relate this equation with the Geodesic equation of motion,
%
%\bea{}
%\frac{d^2\hat{x}^\Lambda}{dt^2}=-\Gamma^\Lambda_{00}
%\label{25}
%\eea
%%\bea{}
%%\frac{d^2x^\Lambda}{dt^2}=-\frac{1}{2}g^{\Lambda\Delta}\left[\frac{\partial g_{0\Lambda}}{\partial x^0}+\frac{\partial g_{0\Lambda}}{\partial x^0}-\frac{\partial g_{00}}{\partial x^\Lambda}\right]\nonumber
%%\eea
%\bea{}
%\frac{d^2\hat{x}^\Lambda}{d\hat{t}^2}=\frac{1}{2}g^{\Lambda\Gamma}\frac{\partial g_{00}}{\partial \hat{x}^\Lambda}\nonumber
%\eea
%%Applying conformal transformation
%%\bea{}
%%\hat{g}_{\Lambda\Gamma}=\phi^{\frac{4}{3n-1}}\eta_{LG}\nonumber
%%\eea
%%\bea{}
%%\hat{g}^{\Lambda\Gamma}=\phi^{\frac{-4}{3n-1}}\eta^{LG}\nonumber
%%\eea
%%\bea{}
%%\frac{d^2x^\Lambda}{dt^2}=\frac{1}{2}g^{SG}\phi^\frac{-4}{5}({\partial_\Lambda}g_{00}\phi^\frac{4}{5})\nonumber
%%\eea
%%Using,
%%\bea{}
%%\hat{\partial}_\Lambda&=&\phi^{\frac{4}{5}}\partial_L\nonumber
%%\eea
%%\bea{}
%%\frac{d^2x^\Lambda}{dt^2}=\frac{1}{2}(\partial_L\phi^\frac{4}{5}\eta_{00})\nonumber
%%\eea
%%\bea{}
%%\frac{d^2x^\Lambda}{dt^2}=-\frac{1}{2}(\partial_L\phi^\frac{4}{5})\nonumber
%%\eea
%%\bea{}
%%M\frac{d^2x^\Lambda}{dt^2}=-\frac{1}{2}M(\partial_L\phi^\frac{4}{5})\nonumber
%%\eea
%
%This equation after using conformal transformation can be written as
%
%\bea{}
%M\frac{d^2\hat{x}^\Lambda}{d\hat{t}^2}=-\frac{1}{2}M(\nabla\phi^\frac{4}{5})\nonumber
%\eea
%
%which is just like the equation of motion in "Newtonian gravity".% We arrive at this result as one arrives at newton gravity by using weak equivalence principle in newton cartan theory.
\\
To conclude the above two sections, one must note the following:
\begin{itemize}
  \item The equations \eqref{7}-\eqref{10} of two entangled particles have been rephrased in nonlocal theory. Though the developed geometric theory is intimately related to local theory of general relativity, the non-locality here is attributed to the extra dimensions. The symmetrization condition too accounts for nonlocal interactions in the geometric theory. This requires the coordinates of each particle to be identical so in order to change one particle's coordinates the changes need to be made simultaneously for the other particle. Please note that, although the geometric translation of quantum equations follows from the Einstein-Hilbert action, it is still different from GR in the context that space and time are not on same footing and also gravity works in $4D$ spacetime not in $1+6$ dimensional configuration space.
  \item The dual to equations of two particle entanglement in geometric theory works with $1+6$ dimensions only. Each particle resides in its own reference frame with three dimensions of space and a common temporal dimension.
  \item Eq.\eqref{16} and Eqs.\eqref{22}-\eqref{24} are merely the translation of quantum equations Eqs.\eqref{7}-\eqref{10} in a geometric language. A set of matching conditions is defined to connect the quantum equations with their geometrical counterparts. These conditions are chosen so as to give the best connecting link between the two theories.
 These conditions connect the quantum phase $S$ with the Hamilton principle function $S_H$, the amplitude of pilot wave $P$ with the conformal function of the metric $\phi$ and the mass $m_j$ with the mass $\hat{M}_G$. The coupling constant of the geometric theory is
\bea{}
k = -\frac{24}{5}.\frac{2\hat{M}_G}{\hbar^2}\nonumber
\eea
Further, please note that the particles are strongly coupled to the applied external magnetic field that makes them experience torque, but coupling on the geometric side of this duality is weak ($k<1$). Hence, by means of the matching conditions (to switch between geometry and physics of the two particle entanglement), we arrive at a strong-weak duality. Strong on the physics side and weak on the geometric side.
\end{itemize}
\section{ENTANGLEMENT equations in terms of FINSLER geometry}

Finsler geometry gives insight into a novel approach to discuss the dynamics and geometry of matter fields. GR works with a geometric background furnished with a $4D$ Lorentzian manifold to put field theories into causal geometrical interpretation. One can however expand this geometric background to a non-metric, general length measure background Finsler spacetime \cite{29,30,31}. The dynamics thus described is compatible with GR.

The Einstein-Hilbert action including matter field interaction is given by
\bea{}
S[g,\phi_i]=\int\limits_M d^4x\sqrt{|g|}\left(R+kL_M[g,\phi_i]\right)
\eea

In the limit when we replace

\bea{}
\sqrt{|g|}=\sqrt{|g|}\sqrt{|h|}
\label{26}
\eea
\bea{}
R=R_{ab}
\label{27}
\eea
The Einstein-Hilbert action could then be written for Finsler space in terms of a general length measure $F$ (Finsler function) over a tangent bundle $TM$. We consider the sphere $S_P$ to be fibred over each point of the $4D$ spacetime manifold $M$, in the tangent space $T_PM$
\bea{}
S_P=\left[y\in T_PM\mid\sqrt{F_{\mid P}(y,y)}=1\right]\nonumber
\eea
The Einstein-Hilbert action can then be written as an action on the sphere bundle $\sum$ which is subset of tangent bundle $TM$ obtained by union over all points as
\bea{}
S_P\subset T_PM\nonumber
\eea
Introducing the notion
\bea{}
(x^a,\theta^\alpha),\,\,\,\,\,\,\ a=0,1,2,3 ,\,\,\,\,\,\ \alpha=1,2,3\nonumber
\eea
The resulting Einstein-Hilbert action in Finsler space for sphere bundle becomes \cite{30}
\bea{}
S[F,\phi_i]=\int\limits_\Sigma d^4x d^3\theta\sqrt{|g||h|}\left(\hat{R}_{ab}+k\hat{L}_M[g,\phi_i]\right)
\label{28}
\eea
where $F$ is the Finsler function. Also note that the coupling constant arising in Eq.\eqref{28} is different from that of the GR coupling constant.

using
\bea{}
R_{ab}=g^{ab}R
\eea
%\bea{}
%\hat{R}_{ab}=\phi^{-\frac{9}{5}}\hat{g}^{ab}\left[\frac{-4(4+3-1)}{4+3-2}\Delta\phi^{-4/5}+R\phi^{-4/5}\right]\nonumber
%\eea
%\bea{}
%\hat{R}_{ab}=\phi^{-\frac{9}{5}}\hat{g}^{ab}\left[\frac{-24}{5}\nabla^2\phi^2\phi^{-14/5}+R\phi^2\phi^{-14/5}\right]\nonumber
%\eea
%\bea{}
%\hat{R}_{ab}=\phi^{-27/5}g^{ab}\left[\frac{-24}{5}\partial^L\phi\partial_L\phi+R\phi^2\right]
%\eea
%Using equation (29) in (28),
and the conformal transformation,
\bea{}
{\hat{g}_{\Lambda\Gamma}=\phi^{4/5}(x_j,\Theta)\eta_{LG}}
\eea
{The inverse of the metric is given by}
\bea{}
{\hat{g}^{\Lambda\Gamma}=\phi^{-4/5}(x_j,\Theta)\eta^{LG}}
\eea
 The particle action in terms of $\phi$ and Finsler function $F$ becomes

\bea{}
S[F,\phi_i]=\int\limits_\Sigma d^4x d^3\theta\sqrt{|g||h|}[\hat{g}^{ab}\phi^{-23/5}\times\nonumber
\\
\left(\frac{-24}{5}\partial^L\phi\partial_L\phi+R\phi^2\right)+\nonumber
\\
k\phi^2\phi^{-14/5}\left(L_M-\frac{\mu_j.B(x_j)\textbf{S}_j}{S_j}\right)]
\eea

when we choose $R=0$,

\bea{}
S[F,\phi_i]=\int\limits_\Sigma d^4x d^3\theta\sqrt{|g||h|}\phi^{-14/5}\times~~~~~~~~~~~\nonumber
\\
\left[-\phi^{-13/5}\frac{24}{5}\partial^L\phi\partial_L\phi+k\phi^2\left(L_M-\frac{\mu_j.B(x_j)\textbf{S}_j}{S_j}\right)\right]
\eea

\bea{}
\Rightarrow-\phi^{-13/5}\frac{24}{5}\frac{\partial^L\partial_L\phi}{\phi}=k\left(L_M-\frac{\mu_j.B(x_j)\textbf{S}_j}{S_j}\right)
\label{15}
\eea
The matter Lagrangian $L_M$ is given by
%\bea{}
%L_M=\frac{2\hat{\pi}^\Lambda\hat{\pi}_\Lambda}{2\hat{M}_G}+\frac{\partial S_H(\Theta,\vec{x}_j)}{\partial \Theta}\nonumber
%\eea
\bea{}
L_M=\frac{2(\hat{\partial}^\Lambda S_H-ie_j\hat{A}^\Lambda(x_j))(\hat{\partial}_\Lambda S_H-ie_j\hat{A}_\Lambda(x_j))}{2\hat{M}_G}+~~~~\nonumber
\\
\frac{\partial S_H}{\partial \Theta}~~~~~~~~~~~~
\eea
The equation of motion with this Lagrangian then gives
\bea{}
-{2\hat{M}_G}\phi^{-13/5}\frac{24}{k(5)}\frac{\partial^L\partial_L\phi}{\phi}=2(\hat{\partial}^\Lambda S_H)(\hat{\partial}_\Lambda S_H)-2e^2_j\hat{A}^{\Lambda2}(x_j)\nonumber
\\
-{2\hat{M}_G}\frac{\mu_j.B(x_j)\textbf{S}_j}{S_j}+{2\hat{M}_G}\frac{\partial S_H}{\partial\Theta}~~~~~~~~~~~~~~
\label{29}
 \eea
With the following matching conditions
\bea{}
k=-\phi^{-13/5}\frac{24}{5}.\frac{2\hat{M}_G}{\hbar^2}
\eea
\bea{}
\phi(\vec{x}_j,\Theta)=P(\vec{x}_j,t)
\eea
\bea{}
S_H(\vec{x}_j,\Theta)=S(\vec{x}_j,t)
\eea
\bea{}
m_j=\hat{M}_G
\eea
\bea{}
\frac{\partial}{\partial t}=\frac{\partial}{\partial \Theta}
\eea
where
\bea{}
\partial \Theta=\partial \theta_\alpha,\,\,\,\,\,\,\ \alpha=1,2,3
\eea

\eqref{29} is identical to \eqref{7}.

\subsection{Geometric dual to the second equation:}

As before, using the conservation of energy-momentum tensor and following the same procedure as done for the previous dual (see appendix C), we obtain

\bea{}
\frac{(\hat{\partial}_\Lambda S_H)\nabla_\Lambda(\hat{\partial}^\Lambda S_H)}{\hat{M}_G}+\nabla_\Lambda\frac{\partial S_H}{\partial\Theta}=0
\label{k}
\eea

Using the definition of Levi-Civita connection \eqref{j} in \eqref{k}, we obtain

\bea{}
\frac{(\hat{\partial}_\Lambda S_H)\nabla_\Lambda(\hat{\partial}^\Lambda S_H)}{\hat{M}_G}=\phi^{\frac{-14}{5}}\partial_L(\phi^2\partial^L S_H)=0
\label{33}
\eea
and
\bea{}
\nabla_\Lambda\frac{\partial S_H}{\partial\Theta}=\frac{\partial}{\partial\Theta}(\partial_L S_H)
\label{34}
\eea

One obtains from \eqref{33} and \eqref{34}
\bea{}
\left(\frac{\partial_L\left[\phi^2(\partial^LS_H)\right]}{\hat{M}_G}+\frac{\partial\phi^2}{\partial\Theta}\right)=0
\label{35}
\eea

Eq.\eqref{35} with the defined matching conditions is identical to Eq.\eqref{8}
\subsection{Geometric dual to the third equation:}
The particle trajectories are governed by

\bea{}
\hat{\pi}^\Lambda=(\hat{\partial}^\Lambda S_H-ie_j\hat{A}^\Lambda(x_j))
\label{36}
\eea
Eq.\eqref{36} with the defined matching conditions is identical to Eq.\eqref{9}.

\subsection{Geometric dual to the equation of motion:}

The Finsler geometric dual to the trajectory equation of motion Eq.\eqref{10} is obtained in a similar way as done for the earlier case of configuration space geometry. (see appendix D), one finds

\bea{}
\hat{M}_G\frac{d^2\hat{x}^\Lambda}{d\hat{\Theta}^2}=\hat{\partial}^\Lambda[Q
+\frac{\mu_j.B(x_j)\textbf{S}_j}{S_j}
\nonumber\\
-\frac{ieS_H\hat{F}_{\Lambda}^*}{\hat{M}_G}-\frac{\partial S_H}{\partial\Theta}
]
\eea
This equation describes the dynamics of particle in Finsler spacetime and is identical to Eq.\eqref{10} with given matching conditions.
\\
To conclude this section, please note the following:
\begin{itemize}
  \item For geometry developed over configuration space, nonlocality is encoded into the theory by means of quantum potential, symmetrization condition and extra dimensions. In Finslerian model, however, quantum potential is responsible for nonlocality. This potential effects the particles in such a way that it is not possible to isolate one particle from the other, asserting that the two particles are entangled.

  \item Finsler geometry is developed in seven dimensions over a Finslerian manifold which is fibered over by a unit sphere at each point. One must note that, we could restore gravity from Finsler spacetime if the general length measure (Finsler function $F$) is identical to the metric length or by means of equations \eqref{26}, \eqref{27}.

  \item The Finsler geometry is connected with the physical equations of two particle entanglement by defining an appropriate set of matching conditions. These conditions connect the quantum phase $S$ with the Hamilton principle function $S_H$, the amplitude of pilot wave $P$ with the conformal function of the metric $\phi$, the mass $m_j$ with the mass $\hat{M}_G$ and the time coordinate $t$ with polar angle $\Theta$. The coupling required to match the physics side of the duality with (Finsler) geometric side is given by
        \bea{}
k=-\phi^{-13/5}\frac{24}{5}.\frac{2\hat{M}_G}{\hbar^2}
\label{j}
\eea
The coupling constant arising in the (Finsler) action in Eq.\eqref{28}, is the following
\bea{}
k=\frac{4\pi G}{c^4 }.\frac{1}{y^a y^b}
\label{k}
\eea
Note that the coupling constant in Eq.\eqref{j} includes $\phi^{-13/5}$, %\textcolor[rgb]{1.00,0.00,0.00}{defined over sphere bundle $S\subset T_PM$. And the tangent bundle $TM$ has $4D$-spacetime as its base manifold}.
This function accounts for the matter-field interaction and is found to be $\phi^{-13/5}=-\frac{8\pi G}{c^4 }.\frac{1}{y^a y^b}\frac{5}{24}.\frac{\hbar^2}{4\hat{M}_G}$. The term $\frac{8\pi G}{c^4 }$ is the Einstein's constant. Thus, the particles being very light (e.g., electrons) are weakly affected by the gravitational field. Therefore, one can conclude that the strongly coupled quantum equations have a dual geometric description in terms of weakly coupled equations in Finsler framework.
\end{itemize}

\section{SUMMARY}

We studied the geometric duality of the equations of two entangled particles with configuration space and Finsler space respectively. The physics side of the duality constitutes a set of four quantum equations for two entangled particles with spin, in the pilot-wave limit. We presented two types of geometric dualities for this set of four equations. The first geometric description follows from the action \eqref{12} where the particles move along geodesics over a $1+6$ dimensional configuration space. The second geometric description follows from the Einstein-Hilbert extended Finsler action \eqref{28} over a seven dimensional manifold.

The constant $\kappa$ specifying the matter field interaction in both geometric theories is  different and that's because, the Finsler gravity action includes Ricci tensor in contrast to the configuration space one. This makes the Hamilton-Jacobi equation \eqref{3} slightly different in the two geometric formulations while the other duals appears to be merely a reformulation of \eqref{8}-\eqref{10} in two different geometries. In each case a suitable set of matching conditions is defined to connect the geometric side (configuration space and Finsler space respectively) with the equations of quantum entanglement.

The duality presented in this paper is such that one could translate from a strongly coupled quantum theory in the pilot-wave limit, to weakly coupled geometric theories (either $(1+6)D$ configuration space or Finsler space). The (two entangled) particles have strong interaction with the external magnetic field that causes the particles to experience torque. One can deduce from the two (geometric) couplings
\bea{}
k=-\frac{24}{5}.\frac{2\hat{M}_G}{\hbar^2}
\eea
and
\bea{}
k=-\phi^{-13/5}\frac{24}{5}.\frac{2\hat{M}_G}{\hbar^2}
\eea
that the strongly coupled quantum equations have a correspondence with weakly coupled configuration space and Finsler space geometric theories.

\section{{Appendix}}
\subsection{First Equation}
Starting from Eq.\eqref{06}
%\bea{}
%\psi(x_1,x_2,t)=P(x_1,x_2,t)exp\left(\frac{iS(x_1,x_2,t)}{\hbar}\right)
%\eea
%We work in the limit of absolute "time", designated as $t$. In this limit, space and time are not on equal footing so the
%spatial and temporal derivatives will be treated differently

\bea{}
\sum\limits_j\left(\frac{\hbar^2}{2m_j}\partial^K_j\partial_{jK}+\frac{e^2_j}{2m_j}A_j^{K2}(x_j)+\frac{\mu_j.B(x_j)\textbf{S}_j}{S_j}+i\hbar\partial_0\right)\nonumber
\\
\times P(x_1,x_2,t)\textmd{exp}\left(\frac{iS(x_1,x_2,t)}{\hbar}\right)=0~~~~~~~~~~
\eea

%\bea{}
%\sum\limits_j[\frac{\hbar^2}{2m_j}\partial^K_j\partial_{jK}Pe^\frac{iS}{\hbar}+\frac{i\hbar^2}{2m_j\hbar}P\partial^K_j(\partial_{jK}S)e^\frac{iS}{\hbar}+\nonumber
%\\\frac{e^2_j}{2m_j}A_j^{K2}(x_j)Pe^\frac{iS}{\hbar}
%+\frac{\mu_j.B(x_j)\textbf{S}_j}{S_j}Pe^\frac{iS}{\hbar}-P\partial_0S]=0\nonumber
%\eea
leads to the following
\bea{}
\sum\limits_j[\hbar^2\partial^K_j\partial_{jK}P+\frac{i\hbar^2}{\hbar}P\partial^K_j(\partial_{jK}S)\frac{iS}{\hbar}+e^2_jA_j^{K2}(x_j)P\nonumber\\
+{2m_j}\frac{\mu_j.B(x_j)\textbf{S}_j}{S_j}P-{2m_j}P\partial_0S]=0~~~~~~~~~~~~~~~
\eea
On the average for large $t$, $\partial P/\partial t=0$. Taking real part of the above equation after using Taylor series, we get
\bea{}
{2m_j}\sum\limits_j\frac{\hbar^2}{{2m_j}}\frac{\partial^K_j\partial_{jK}P}{P}=\sum\limits_j[(\partial^K_jS)(\partial_{jK}S)-e^2_jA_j^{K2}(x_j)
\nonumber\\-{2m_j}\frac{\mu_j.B(x_j)\textbf{S}_j}{S_j}+{2m_j}\partial_0S]~~~~~~~~~~~~~
\label{b}
\eea
Note that the term $\sum\limits_j\frac{\hbar^2}{{2m_j}}\frac{\partial^K_j\partial_{jK}P}{P}$ on the left hand side of \eqref{b} is the formal definition of quantum potential $Q$. So,
\bea{}
{2m_j}Q(x_1,x_2,t)=\sum\limits_j[(\partial^K_jS)(\partial_{jK}S)-e^2_jA_j^{K2}(x_j)\nonumber
\\
-{2m_j}\frac{\mu_j.B(x_j)\textbf{S}_j}{S_j}+{2m_j}\partial_0S]
\eea
\subsection{Trajectory Equation of Motion}
Using the definition of momenta
\bea{}
\frac{dx}{dt}=\frac{(\partial_j^KS)-ie_jA_j^K(x_j)}{m_j}
\eea
and using the identity
\bea{}
\frac{d}{dt}=\sum\limits_l\frac{d}{dx_l^N}\frac{dx_l^N}{dt}
\eea
one could derive the equation of motion for two entangled particles as
\bea{}
\frac{d^2x_j^K}{dt^2}=\sum\limits_l\frac{d}{dx_l^N}\frac{dx_l^N}{dt}\frac{(\partial_j^KS)-ie_jA_j^K(x_j)}{m_j}
\eea

%which gives
%
%\bea{}
%m_j^2\frac{d^2x_j^K}{dt^2}=\sum\limits_l\partial_{lN}\pi_l^N\pi_j^K\nonumber
%\eea
%\bea{}
%m_j^2\frac{d^2x^L}{dt^2}=\partial_N\pi^N\pi^L\nonumber
%\eea
%\bea{}
%\frac{d^2x_j^K}{dt^2}=\sum\limits_l\partial_{lN}\left[
%\frac{(\partial_l^NS)-ie_lA_l^N(x_l)}{m_j}\right]\nonumber
%\\
%\times\left[\frac{(\partial_j^KS)-ie_jA_j^K(x_j)}{m_j}\right]\nonumber
%\eea
%\bea{}
%m_j\frac{d^2x_j^K}{dt^2}=\sum\limits_l\partial_{lN}\left[
%\frac{(\partial_l^NS)-ie_lA_l^N(x_l)}{m_j}\right]\left[(\partial_j^KS)-ie_jA_j^K(x_j)\right]\nonumber
%\eea

%\bea{}
%m_j^2\frac{d^2x_j^K}{dt^2}=\sum\limits_l\left(\partial_{lN}[(\partial_l^NS)(\partial_j^KS)-(\partial_j^KS)ie_lA_l^N(x_l)
%-(\partial_l^NS)ie_jA_j^K(x_j)+ie_lA_l^N(x_l)ie_jA_j^K(x_j)]\right)\nonumber
%\eea
%\bea{}
%m_j^2\frac{d^2x_j^K}{dt^2}=\sum\limits_l[\partial_{lN}[(\partial_l^NS)(\partial_j^KS)-ieS(\partial_j^K A_l^N+\partial_l^N A_j^K)\nonumber\\
%+ie_lA_l^N(x_l)ie_jA_j^K(x_j)]]~~~~~~~
%\eea
After some algebra, one obtains
\bea{}
m_j^2\frac{d^2x_j^K}{dt^2}=\sum\limits_l(\partial_{lN}[(\partial_l^NS)(\partial_j^KS)-ieSF^*\nonumber
\\+ie_lA_l^N(x_l)ie_jA_j^K(x_j)])
\eea
where, the field strength tensor is given by
\bea{}
F^*=\partial_j^K A_l^N+\partial_l^N A_j^K
\eea
%For $N=K=L$
Using single index notation where $\partial_j^K,\partial_l^N\rightarrow\partial^L$
\bea{}
m_j^2\frac{d^2x^L}{dt^2}=\left(\partial_{L}[(\partial^LS)(\partial_LS)-ieSF^*-e^2A^{L2}(x_l)]\right)
\eea

Since,
%\bea{}
%Q(x_1,x_2,t)=[\frac{(\partial^LS)(\partial_{L}S)}{2m_j}-\frac{e^2_j}{2m_j}A^{L2}(x_j)-\frac{\mu_j.B(x_j)\textbf{S}_j}{S_j}\nonumber\\
%+\partial_0S]\nonumber
%\eea
\bea{}
2m_j\left[ Q+\frac{\mu_j.B(x_j)\textbf{S}_j}{S_j}-\partial_0S\right]=(\partial^L S)(\partial_L S)-e^2_jA^{L2}(x_j)~~
\eea
%\bea{}
%m_j^2\frac{d^2x_j^L}{dt^2}=\partial^{L}\left[2m_j \left[Q+\frac{\mu_j.B(x_j)\textbf{S}_j}{S_j}-\partial_0S\right]-ieSF^*\right]
%\eea
%For instantaneous information processing, we may choose $\partial_0S=0$.
So, the trajectory equation of motion becomes
\bea{}
m_j\frac{d^2x^L}{dt^2}=\partial_{L}\left[2 Q+\frac{2\mu_j.B(x_j)\textbf{S}_j}{S_j}-\frac{ieSF^*}{m_j}-2\partial_0S\right]~~
\eea
\subsection{Geometric dual to 2nd Equation}
The stress-energy tensor is given by
\bea{}
T^{\Lambda\Delta}=2\frac{\delta L_M}{\delta g_{\Lambda\Delta}}+g^{\Lambda\Delta}L_M
\eea

%\bea{}
%T^{\Lambda\Delta}=2\frac{\delta}{\delta g_{\Lambda\Delta}}[g_{\Lambda\Delta}\frac{(\hat{\partial}^\Lambda S_H)(\hat{\partial}^\Delta S_H)}{\hat{M}_G}+\partial_0S_H-\frac{e_j^2\hat{A}^{\Lambda2}}{\hat{M}_G}\nonumber\\
%-\frac{\mu_j.B(x_j)\textbf{S}_j}{S_j}]+
%\frac{g^{\Lambda\Delta}(\hat{\partial}^\Lambda S_H)(\hat{\partial}_\Lambda S_H)}{\hat{M}_G}+g^{\Lambda\Delta}\partial_0S_H\nonumber\\-g^{\Lambda\Delta}\frac{e_j^2\hat{A}^{\Lambda2}}{\hat{M}_G}-\frac{\mu_j.B(x_j)\textbf{S}_j}{S_j}\nonumber
%\eea
Substituting for matter Lagrangian one finds
\bea{}
T^{\Lambda\Delta}=\frac{2(\hat{\partial}^\Lambda S_H)(\hat{\partial}^\Delta S_H)}{\hat{M}_G}+~~~~~~~~~~~~~~~~~~~~~~~~~~~~~\nonumber
\\
g^{\Lambda\Delta}\left(\frac{(\hat{\partial}^\Lambda S_H)(\hat{\partial}_\Lambda S_H)}{\hat{M}_G}+
\partial_0S_H-\frac{e_j^2\hat{A}^{\Lambda2}}{\hat{M}_G}\right)~~~
\eea

where $\partial_0=\partial/\partial t$ and since the stress energy tensor is covariantly conserved $\nabla_\Lambda T^{\Lambda\Delta}=0$, so
\bea{}
\frac{2\nabla_\Lambda(\hat{\partial}^\Lambda S_H)(\hat{\partial}^\Delta S_H)}{\hat{M}_G}+\frac{2(\hat{\partial}^\Lambda S_H)g^{\Lambda\Delta}
\nabla_\Lambda(\hat{\partial}_\Lambda S_H)}{\hat{M}_G}+~~~~~~~~~~~~~~~~~~~~~~~~~~~~~~~~~~~~~~~~~~~~~\nonumber\\\frac{\nabla_\Lambda g^{\Lambda\Delta}(\hat{\partial}_\Lambda S_H)(\hat{\partial}^\Lambda S_H)}{\hat{M}_G}+\frac{g^{\Lambda\Delta}\nabla_\Lambda(\hat{\partial}_\Lambda S_H)(\hat{\partial}^\Lambda S_H)}{\hat{M}_G}+
~~~~~~~~~~~~~~~~~~~~~~~~~~~~~~~~~~~~~~~~~~~\nonumber\\\frac{g^{\Lambda\Delta}(\hat{\partial}_\Lambda S_H)\nabla_\Lambda(\hat{\partial}^\Lambda S_H)}{\hat{M}_G}+\nabla_\Lambda g^{\Lambda\Delta}(\partial_0 S_H)~~~~~~~~~~~~~~~~~~~~~~~~~~~~~~~~~~~~~~~~~~~~~
\nonumber
\\+g^{\Lambda\Delta}\nabla_\Lambda(\partial_0 S_H)=0~~~~~~~~~~~~~~~~~~~~~~~~~~~~~~~~~~~~~~~~~~~~~~~~\
\eea
The covariant derivative of the metric is taken to be zero, $\nabla_\Lambda g^{\Lambda\Delta}=0$. So that one finds
\bea{}
\frac{2(\hat{\partial}^\Delta S_H)\nabla_\Lambda(\hat{\partial}^\Lambda S_H)}{\hat{M}_G}+\frac{2(\hat{\partial}^\Lambda S_H)\nabla^\Delta(\hat{\partial}_\Lambda S_H)}{\hat{M}_G}
\nonumber\\
+\frac{(\hat{\partial}^\Lambda S_H)\nabla^\Delta(\hat{\partial}_\Lambda S_H)}{\hat{M}_G}+\frac{g^{\Lambda\Delta}(\hat{\partial}_\Lambda S_H)\nabla_\Lambda(\hat{\partial}^\Lambda S_H)}{\hat{M}_G}\nonumber\\~~~~~~~~~
+g^{\Lambda\Delta}\nabla_\Lambda(\partial_0S_H)=0~~~~~~~~~~~~~~~~~~~~
\eea
From this, we can write

\bea{}
\frac{(\hat{\partial}^\Delta S_H)\nabla_\Lambda(\hat{\partial}^\Lambda S_H)}{\hat{M}_G}=0
\label{c}
\eea

\bea{}
\frac{(\hat{\partial}^\Lambda S_H)\nabla^\Delta(\hat{\partial}_\Lambda S_H)}{\hat{M}_G}=0
\label{d}
\eea
\bea{}
\frac{(\hat{\partial}_\Lambda S_H)\nabla_\Lambda(\hat{\partial}^\Lambda S_H)}{\hat{M}_G}+\nabla_\Lambda(\partial_0 S_H)=0
\label{e}
\eea
%With this, the relation in \eqref{e} reads:
%\\
%
%For the first term in \eqref{e}
%
%\bea{}
%\nabla_\Lambda(\hat{\partial}^\Lambda S_H)=\partial_\Lambda(\partial^\Lambda S_H)+~~~~~~~~~~~~~~~~~~~~~~~~~\nonumber
%\\
%\frac{1}{2}g^{\Sigma\Xi}[\partial_\Lambda g_{\Delta\Xi}+\partial_\Delta g_{\Xi\Lambda}-\partial_\Xi g_{\Lambda\Delta}][\partial^\Lambda S_H]=0
%\eea
%\bea{}
%\phi^{\frac{-14}{5}}\partial_L(\phi^2(\partial^LS_H))=0
%\eea
%
%For the second term in \eqref{e}
%
%\bea{}
%\nabla_\Lambda(\partial_0S_H)=\partial_\Lambda(\partial_0S_H)
%\eea
%\bea{}
%\nabla_\Lambda(\partial_0S_H)=\partial_0(\partial_LS_H)
%\eea
%\\
%
%From \eqref{e}, with first term and second term, we get
%
%\bea{}
%\left(\frac{\partial_L\left[\phi^2(\partial^LS_H)\right]}{\hat{M}_G}+\partial_0(\phi^2)\right)=0
%\eea
\subsection{Dual to trajectory equation of motion}
The total derivative is

\bea{}
\frac{d}{d\hat{s}}&=&\frac{d\hat{x}^\Lambda}{d\hat{s}}\hat{\partial}_\Lambda
\eea
Using conformal transformation, one obtains
\bea{}
\frac{d}{d\hat{s}}&=&\phi^{\frac{4}{5}}\frac{dx^L}{ds}\partial_
\\
\frac{d}{d\hat{s}}&=&\phi^{\frac{4}{5}}\frac{d}{ds}
\eea

Applying this relation to momenta
\bea{}
\frac{d^2\hat{x}^\Lambda}{d\hat{s}^2}=\frac{d}{d\hat{s}}\frac{\left(\hat{\partial}^\Lambda S_H-ie_j\hat{A}^\Lambda(x_j)\right)}{\hat{M}_G}
\eea

and using the identity

\bea{}
\frac{d}{d\hat{s}}=\frac{d}{d\hat{x}^\Lambda}\frac{{d\hat{x}^\Lambda}}{d\hat{s}}
\eea
\bea{}
\frac{d}{d\hat{s}}=\hat{\partial}_\Lambda\frac{{d\hat{x}^\Lambda}}{d\hat{s}}
\eea
one could derive trajectory equation of motion as
\bea{}
\frac{{d^2\hat{x}^\Lambda}}{d\hat{s^2}}=\hat{\partial}_\Delta\frac{{d\hat{x}^\Delta}}{d\hat{s}}\frac{\left(\hat{\partial}^\Lambda S_H-ie_j\hat{A}^\Lambda(x_j)\right)}{\hat{M}_G}
\label{m}
\eea
Using the definition of momentum and after doing some algebra, \eqref{m} yields
%\bea{}
%\hat{M}_G^2\frac{d^2\hat{x}^\Lambda}{d\hat{s}^2}=\hat{\partial}_\Delta\hat{\pi}^\Delta\hat{\pi}^\Lambda\nonumber
%\eea
%\bea{}
%\frac{d^2x_j^K}{dt^2}=\sum\limits_l\partial_{lN}\left[
%\frac{(\partial_l^NS)-ie_lA_l^N(x_l)}{m_j}\right]\left[\frac{(\partial_j^KS)-ie_jA_j^K(x_j)}{m_j}\right]\nonumber
%\eea
%\bea{}
%m_j\frac{d^2x_j^K}{dt^2}=\sum\limits_l\partial_{lN}\left[
%\frac{(\partial_l^NS)-ie_lA_l^N(x_l)}{m_j}\right]\left[(\partial_j^KS)-ie_jA_j^K(x_j)\right]\nonumber
%\eea

%\bea{}
%M_G^2\frac{d^2\hat{x}^\Lambda}{d\hat{s}^2}=\hat{\partial}_\Delta[(\hat{\partial}^\Delta S_H)(\hat{\partial}^\Lambda S_H)-(\hat{\partial}^\Delta S_H)ie_j\hat{A}^\Lambda(x_j)\nonumber
%\\
%-(\hat{\partial}^\Lambda S_H)ie_j\hat{A}^\Delta(x_j)+ie_j\hat{A}^\Delta(x_j)ie_j\hat{A}^\Lambda(x_j)])\nonumber
%\eea
%\bea{}
%M_G^2\frac{d^2\hat{x}^\Lambda}{d\hat{s}^2}=\hat{\partial}_\Delta[(\hat{\partial}^\Delta S_H)(\hat{\partial}^\Lambda S_H)-ie_jS_H(\hat{\partial}^\Delta \hat{A}^\Lambda+
%\nonumber \\
%\hat{\partial}^\Lambda \hat{A}^\Delta)+ie_j\hat{A}^\Delta(x_j)ie_j\hat{A}^\Lambda(x_j)]~~~~
%\eea

\bea{}
\hat{M}_G^2\frac{d^2\hat{x}^\Lambda}{d\hat{s}^2}=\hat{\partial}_\Delta[(\hat{\partial}^\Delta S_H)(\hat{\partial}^\Lambda S_H)-ie_jS_H\hat{F}^{\Delta*}+
\nonumber\\
ie_j\hat{A}^\Delta(x_j)ie_j\hat{A}^\Lambda(x_j)]
\label{n}
\eea
with the field strength tensor given by
\bea{}
\hat{F}^{\Delta*}=(\hat{\partial}^\Delta \hat{A}^\Lambda+
\hat{\partial}^\Lambda \hat{A}^\Delta)
\eea
%\bea{}
%M_G^2\frac{d^2\hat{x}^\Lambda}{d\hat{s}^2}=\hat{\partial}_\Delta[g^{\Lambda\Delta}(\hat{\partial}_\Lambda S_H)(\hat{\partial}^\Lambda S_H)-ie_jS_Hg^{\Lambda\Delta}\hat{F}_{\Lambda}^*
%\nonumber\\
%-e^2_jg^{\Lambda\Delta}\hat{A}_\Lambda \hat{A}^\Lambda(x_j)]\nonumber
%\eea
Changing the index $\Delta$ in \eqref{n} inside the parenthesis using the metric gives
\bea{}
\hat{M}_G^2\frac{d^2\hat{x}^\Lambda}{d\hat{s}^2}=\hat{\partial}^\Lambda[(\hat{\partial}_\Lambda S_H)(\hat{\partial}^\Lambda S_H)-ie_jS_H\hat{F}_{\Lambda}^*
\nonumber\\
-e^2_j\hat{A}_\Lambda \hat{A}^\Lambda(x_j)]
\label{i}
\eea
Substituting
\bea{}
(\hat{\partial}_\Lambda S_H)(\hat{\partial}^\Lambda S_H)
-e^2_j \hat{A}^{\Lambda 2}(x_j)=M_G[-\frac{24}{k(5)}\frac{\partial^L\partial_L\phi}{\phi}
\nonumber\\
+\frac{\mu_j.B(x_j)\textbf{S}_j}{S_j}-\dot{S}_H]
\eea
in \eqref{i} gives equation of motion.
%\bea{}
%(\hat{\partial}_\Lambda S_H)(\hat{\partial}^\Lambda S_H)
%-e^2_j \hat{A}^{\Lambda 2}(x_j)=M_G[-\frac{24}{k(5)}Q
%\nonumber\\
%+\frac{\mu_j.B(x_j)\textbf{S}_j}{S_j}-\dot{S}_H]\nonumber
%\eea
\bea{}
\hat{M}_G^2\frac{d^2\hat{x}^\Lambda}{d\hat{s}^2}=\hat{\partial}^\Lambda[M_G[-\frac{24}{k(5)}Q
+\frac{\mu_j.B(x_j)\textbf{S}_j}{S_j}-\dot{S}_H]\nonumber\\
-ieS_H\hat{F}_{\Lambda}^*]~~~~~~~~
\eea
\section*{Acknowledgments}

We would like to thank Benjamin Koch for helpful discussions.


\begin{thebibliography}{99}

\expandafter\ifx\csname natexlab\endcsname\relax\def\natexlab#1{#1}\fi
\expandafter\ifx\csname bibnamefont\endcsname\relax
  \def\bibnamefont#1{#1}\fi
\expandafter\ifx\csname bibfnamefont\endcsname\relax
  \def\bibfnamefont#1{#1}\fi
\expandafter\ifx\csname citenamefont\endcsname\relax
  \def\citenamefont#1{#1}\fi
\expandafter\ifx\csname url\endcsname\relax
  \def\url#1{\texttt{#1}}\fi
\expandafter\ifx\csname urlprefix\endcsname\relax\def\urlprefix{URL }\fi
\providecommand{\bibinfo}[2]{#2}
\providecommand{\eprint}[2][]{\url{#2}}


\bibitem{a}I. Tavernelli, Ann. of Phys. , 371 (2016)
\bibitem{6}E. Santamato, J. Math. Phys. 25, 2477 (1984); E. Santamato, Phys. Rev. D 32, 2615 (1985).
\bibitem{11}S. Abraham, P. F. de Cordoba, J. M. Isidro, and J. L. G.
Santander (2008), 0810.2236; S. Abraham, P. F. de Cor-
doba, J. M. Isidro, and J. L. G. Santander (2008),
0810.2356.
\bibitem{12}B. Koch, arXiv:0810.2786 [hep-th].
\bibitem{13}B. Koch, AIP Conf. Proc. 1232, 313 (2010)
doi:10.1063/1.3431507 [arXiv:1004.2879 [hep-th]].
\bibitem{14}B. Koch, AIP Conf. Proc. 1196, 161 (2009)
doi:10.1063/1.3284379 [arXiv:1004.3240 [gr-qc]].
\bibitem{15}P. Nicolini, Phys. Rev. D 82, 044030 (2010)
doi:10.1103/PhysRevD.82.044030 [arXiv:1005.2996 [gr-
qc]].
%\bibitem{16}R. Carroll, arXiv:1007.4744 [math-ph].
\bibitem{17}J. M. Isidro, P. F. de Cordoba, J. M. Rivera-Rebolledo
and J. L. G. Santander, Int. J. Geom. Meth. Mod.
Phys. 8, 621 (2011) doi:10.1142/S0219887811005294
[arXiv:1007.4929 [hep-th]].
\bibitem{18}B. Koch and N. Rojas, Int. J. Geom. Meth. Mod. Phys.
11, 1450029 (2014) doi:10.1142/S0219887814500297
[arXiv:1101.4619 [hep-th]].
\bibitem{19}D. Acosta, P. Fernandez de Cordoba, J. M. Isidro and
J. L. G. Santander, Int. J. Geom. Meth. Mod. Phys.
9, 1250048 (2012) doi:10.1142/S021988781250048X [arX-
iv:1107.1898 [hep-th]].
\bibitem{20}S. H. Mehdipour, Eur. Phys. J. Plus 127, 80
(2012) doi:10.1140/epjp/i2012-12080-4 [arXiv:1111.2468
[gr-qc]].
\bibitem{21}T. S. Bir and P. Vn, Found. Phys. 45, no. 11, 1465 (2015)
doi:10.1007/s10701-015-9920-7 [arXiv:1312.1316 [gr-qc]].
\bibitem{22}H. Nikolic, Bohmian particle trajectories in relativis-
tic bosonic quantum eld theory, Found.Phys.Lett, 17
(2004) 363.
%\bibitem{23}W. Struyve, H. Westman, A new pilot-wave model for
%quantum eld theory, 2006, arXiv: quant-ph/0602229.
\bibitem{24}B. Koch, Int. J. Geom. Meth. Mod. Phys. 10, no. 9,
1320014 (2013). doi:10.1142/S0219887813200144
\bibitem{25}N. Arkani-Hamed and J. Trnka, JHEP 1410, 030
(2014) doi:10.1007/JHEP10(2014)030 [arXiv:1312.2007
[hep-th]].
\bibitem{Alon1}Alon E. Faraggi, M. Matone, The Equivalence Postulate of Quantum Mechanics, Int.J.Mod.Phys. A, 15, 13
(2000), arXiv hep-th/9809127.
\bibitem{Alon2}G. Bertoldi, Alon E. Faraggi, M. Matone, Equivalence
principle, higher dimensional Mobius group and the
hidden antisymmetric tensor of Quantum Mechanics,
Class.Quant.Grav. 17 (2000), arXiv hep-th/9809127.
\bibitem{ADS} J. M. Maldacena, The Large N limit of superconformal field theories and supergravity,
Int.J.Theor.Phys. 38 (1999) 1113-1133.
\bibitem{SQM1} L.\'{A}lvarez-Gaum\'{e}, A note on the Atiyah-Singer index theorem,
J.Phys.A: Mathematical and General, 16, no. 18, pp. 4177-4182, (1983).
\bibitem{SQM2} L.\'{A}lvarez-Gaum\'{e} and E. Witten, Gravitational anomalies,
Nucl.Phys.B, 234, no. 2, pp. 269-330, (1984).
\bibitem{SQM3} J. P. Gauntlett, Low-energy dynamics of $\mathcal{N}=2$ supersymmetric
monopoles, Nucl.Phys.B, 411, no. 2-3, pp. 443-460, (1994).
\bibitem{SQM4} T. J. Hollowood and T. Kingaby, A comment on the $\chi_y$ genus
and supersymmetric quantum mechanics, Phys.Lett.B, 566, no. 3-4, pp. 258-262, (2003).
\bibitem{SQM5} M. A. Wasay, Supersymmetric quantum mechanics and topology, Adv.High Energy Phys. 2016 (2016) 3906746.
\bibitem{SQM6} E. Witten, Constraints on supersymmetry breaking, Nucl.Phys.B, 202, no. 2, pp. 253-316, (1982).
\bibitem{26}D. Bohm, Phys. Rev. 85, 166 (1951).
\bibitem{27}D. Bohm, Phys. Rev. 85, 166 (1951); D. Bohm, Phys.
Rev. 85, 180 (1951);
\bibitem{28}M. A. Wasay, A. Bashir, B. Koch, and A. Ghaffar, Int. J. Geom. Meth. Mod. Phys. 14, 1750149 (2017). doi:10.1142/S0219887817501493
\bibitem{29}C. Pfeifer, The Finsler spacetime framework: backgrounds for physics beyond metric geometry (No. DESY-THESIS-2013-049) (2013).
\bibitem{30}C. Pfeifer, and M. Wohlfarth, In Relativity and Gravitation (pp. 305-308). Springer International Publishing (2014).
\bibitem{31}C. Pfeifer, and M. N. Wohlfarth, Phys. Rev. D, 85(6), 064009 (2012).
\end{thebibliography}
\end{document}